\documentclass[10pt]{article}

\begin{document}

\title{Towards a Post Reductionist Science: \\ The Open Universe}

\author{Stuart Kauffman}

\date{July 8, 2009}

\maketitle

{\par\centering Institute for Biocomplexity and Informatics \\
The University of Calgary \par}
\smallskip 

{\par\centering Signal Processing, Tampere University of Technology \par}
\smallskip 

{\par\centering External Professor, The Santa Fe Institute \par}

\noindent

\begin{abstract}
We have lived with a world view dominated by reductionism.  Yet recently, S. Hawking has written an article entitled "Godel and the End of Physics".  His observations raise the possibility that we should question our foundations. Core to this is reductionism itself.  In turn reductionism finds its roots in Aristotle's model of scientific explanation as deductive inference: All men are mortal, Socrates is a man, therefore Socrates is a mortal. With Newton's laws in differential form, reductionsim snaps into place, for given initial and boundary conditions, integration of those equations is exactly deduction.  Aristotle's 'efficient cause' becomes mathematized as deduction.
In this paper I discuss the reality that deductive inference is not the only way we explain in science. Darwin gave us the Blind Watchmaker, the appearance of design without a designer.   I discuss the role of the opportunity for an adaptation in the biosphere and claim that such an opportunity is a 'blind final cause', not an efficient cause, yet shapes evolution. I also argue that Darwinian exaptations are not describable by sufficient natural law.  Based on an argument of Sir Karl Popper, I claim that no law, or function, f, maps a decoherence process in a Special Relativity setting from a specific space-time slice into its future.  If true this suggests there can be no theory of everything entailing all that happens. I then discuss whether we can view laws as 'enabling constraints' and what they enable. Finally, in place of the weak Anthropic principle in a multiverse, I suggest that we might consider Darwin all the way down. It is not impossible that a single universe has an abiotic natural selection process for laws as enabling constraints and that the single universe that 'wins' is ours.  One possible criterion of winning might be 'most rapid growth of the Adjacent Possible of the universe'.  

\end{abstract}

\bigskip 

\pagebreak

\section*{Introduction}

We have lived in a scientific world view dominated by reductionism for at least 350 years. Reductionism, in S. Weinberg's view, (1) holds that all that unfolds in the universe is logically entailed by the fundamental laws of physics.  In the past thirty five years, doubts as to the full adequacy of reductionism have been increasingly voiced, even in physics. Philip Anderson's "More is Different", (2) and Robert Laughlin's "A Different Universe", (3), are major examples of this doubt Recently, Stephen Hawking has written an article entitled "Godel and the End of Physics", (4), suggesting that no finite set of laws may suffice to describe by entailment the evolution of the universe.  If reductionism proves both profoundly useful but ultimately inadequate, as I think it does, this failure must portend a major change in our scientific world view.  What might follow reductionism as a more fully adequate approach?  I argue that, as this foundation is pulled from under us, it portends a partially lawless open and creative universe of profound new interest.

The heart of what I want to explore begins with this: The very laws of physics may be open to being viewed as \textit{enabling constraints} -  enabling constraint laws \textit{selected by an abiotic natural selection among a set of possible laws to yield our extremely complex universe.} And our single universe, not the multiverse and its attending weak Anthropic principle, may be the 'winning' universe that is enabled by the opportunities afforded by those laws.  In winning, our universe would then have evolved its laws such that the winning universe is ours.  I will discuss initial ideas about what 'winning' might mean below.

By appealing to an abiotic natural selection on a set of laws, this view goes beyond reductionism and explanation purely via logical entailment. As we shall see, Darwin's natural selection goes beyond entailment. We have been taught that science answers only 'how questions'. For example, given Newton's laws and the Newtonian world view, Newton's laws answer how celestial mechanics occurs. But there is no answer to the question, 'Why Newton's Laws'?  Scientific enquiry must stop, on the reductionist view, with the ultimate law, for example Weinberg's Dreams of a Final Theory, (1). \textit{But Darwin's natural selection answers 'why' questions} - why has the vertebrate eye emerged in the evolution of the universe? Because of a sequence of adaptations achieved by Natural Selection, thus achieving what philosopher David Depew calls 'blind teleology',(5). Darwin reaches beyond reductionism because his 'why' question rests on what I shall call 'blind final cause', as is captured in Richard Dawkins famous book, "The Blind Watchmaker", (6). I will argue that the opportunity for an adaptation in the biosphere, or for the universe as a whole, is just such a blind final cause, subsequently achieved by efficient causes. I will suggest and hope to persuade the reader that\textit{ blind final cause is not efficient cause}. This issue will prove central to our discussion.  In sharp contrast, since Descartes and Newton, science has been bound to explain the unfolding of the universe purely in terms of Aristotle's 'efficient cause', \textit{mathematized as logical entailment}.  This assumption is the root of our long faith in reductionism. Given this assumption and the mathematization of efficient cause as entailment, the deductive explanatory, and tautological character, of reductionism is set in place. There is no room for 'lawless creativity' in this world view.  The logical possibility of blind final cause, in the evolution of the biosphere, or even the universe as a whole, renders restriction to our familiar reductionism logically unnecessary, thus goes beyond our familiar reductionism.  Again, reductionism and the consequent faith in deductive entailment yields a universe barren of creativity, a tautological realm entailed by the hoped for theory of everything. In contrast, if 'law' is enabling constraint, and that enablement enables opportunities that can, blindly, be seized by the becoming of the universe in its full becoming, then the universe is open to myriad creativity. The universe is open in ways we have not dreamed in Western science since Descartes. 

This article is organized as follows: In Section 1, I briefly review Aristotle's four causes and his model of scientific explanation in the syllogism. Newton's laws, plus initial and boundary conditions then sets the stage for reductionism, with us today, and leads to Hawking's "Godel and the End of Physics".  I end Section 1 by raising the question whether our sole reliance on efficient cause in science since Newton may be a foundational problem.  In Section 2, I raise the issue of blind final cause in the evolution of the biosphere in the achievement of adaptations that alter the course of the biosphere's evolution, hence that of the universe. Blind final causes are 'opportunities' for adaptations in a selective niche blindly seized by evolution. The selective niche is, itself, not an efficient cause, but a blind final cause of the successful emergence of the adaptation. The selective niche shapes the course of evolution, but is not an efficient cause of that evolution. The adaptation is achieved by efficient causes. The universe is open in ways beyond logical entailment. In Section 3, I describe new grounds to think that the evolution of the biosphere by Darwinian exaptations, or 'preadaptations' is not describable by sufficient natural law, where natural laws are compact descriptions of the regularities of a process.  As we will see, the implications of this evolution by Darwinian preadaptations is that we cannot make probability statements about such evolution since we do not know the sample space of possibilities in what I will call the Adjacent Possible of the biosphere, that in place of sufficient entailing law is a ceaseless creativity, and that the generation of 'information' in the biosphere does not fit Shannon's theory, where Shannon information requires prior knowledge of the ensemble of messages. More the becoming of the biosphere is both partially lawless, yet non-random - a concept we do not yet have in physics.  If these claims are correct, it appears that there can be no entailing Theory of Everything (TOE).  In Section 4, I discuss the issue of whether the co-evolution of the quantum-classical boundary is describable by sufficient natural law and suggest that it is not.  In its place may be an abiotic natural selection blind final cause. These ideas seem to have experimental consequences. In Section 5, I discuss enabling constraints and what they enable. We have not even the beginning of a theory here, but need to develop one.  I will discuss the fact that in the evolution of the biosphere, evolution has itself achieved enabling constraints that have improved the very process of evolution.  If we can view law as enabling constraints, then the biosphere is evolving its own laws such that it evolves better. If so, then it becomes thinkable that the universe has evolved its laws as well. I will sketch an initial approach to what I believe is a non-algorithmic process with an algorithmic set of board games, legal move sets on those board games and the possibilities enabled by those legal rules.  The approach is inadequate, but a start. In Section 6, I discuss our current cosmological conundrum, the apparent fine tuning of the 23 constants of nature, which has led to the suggestion of a multiverse and the weak Anthropic principle.  We are driven to a multiverse hypothesis by reductionism itself. But that reductionism may no longer be all we need. An alternative reductionist hope is that there really is a TOE, and we are left to wonder why this, rather than another TOE, describes the universe.  In place of these familiar theories, I raise the possibility that the laws of physics are enabling constraints that enable a very complex universe, laws that were selected from some set of possible laws early in the history of the universe by blind final cause for a 'winning' persistent and complex universe, and point to some features of physical laws that are puzzling but interesting hints in this context, including various conservation laws.  To pursue the above agenda, if it has merit, will require an entire new body of theory concerning enabling constraints and what they enable.  In the final Section 7, I try to discuss the puzzling role and status of the possible in the origin and history of the universe.  

\section{Aristotle and the Mathematization of \\ Efficient Causes}

Aristotle famously held that there were four causes, formal, final, material and efficient.  In a simple example of a house to be constructed, the formal cause of the house is the blueprint. The material causes of the house are the bricks, mortar, beams, and building material. The final cause of the house is my decision to build the house. The efficient cause is the actual process of its construction.  But Aristotle, as R. Rosen points out in Life Itself (7), also offered a model of scientific explanation in the syllogism: All men are mortal. Socrates is a man. Therefore Socrates is a mortal.  The logical 'force' of this logical entailment may play a later role in our sense that natural efficient cause laws govern rather than describe the unfolding of the universe.

Newton's laws, given as differential equations in a state space with initial and boundary conditions, fulfill Aristotle's form of scientific explanation as entailment, for the integration of the differential equations is precisely deduction.  With Newton, Aristotle's other causes, formal, final, and material, largely recede from science, which takes itself to explain purely in terms of efficient cause, mathematized as logical entailment. 

That this is the base of reductionism is already evident in Laplace's famous claim that a massive computing system, if given the positions and momenta of all the particles in the universe, could, using Newton's laws, predict or retrodict the entire future and past of the universe.  I note four features of Laplace's reductionism: i. All laws are deterministic, now in doubt given quantum mechanics and the Copenhagen interpretation including the Born rule. ii. All that exists ontologically in the universe are particles in motion. iii. All that happens in the universe is describable by sufficient efficient cause laws via deductive entailment. iv. There exists at least one language, here Newton's, to describe all of reality. None of these four claims will remain unchallenged below.  With the addition of fields, quantum mechanics and General Relativity, plus the standard model, we have contemporary reductionist physics and Hawking's doubts, (4).  Indeed, Hawking, in seeing in Godel the potential 'end of physics', does so in terms of the pure sufficiency of a mathematized form of efficient cause law. It is in terms of such mathematized efficient cause laws that he fears an infinite set of such laws.

When such a crisis as Hawking hints arises, one recourse is to doubt the fundamental assumptions we make.  The use of efficient cause as the sole explanatory principle may be just that assumption. I now claim that this assumption is false.  

\section{Blind Final Cause in Biological Evolution}

Darwin may rank as the mind who most changed our world view, for with Darwin we are given, in Richard Dawkin's fine phrase, "the Blind Watchmaker", that is, the emergence in the biosphere of the appearance of design without a designer - the teleonomy of which J. Monod speaks so eloquently in Chance and Necessity,(8).  In this 150th year since Darwin's Origin of Species, we are still grappling with the implications of his central idea.  Philosopher David Depew at a recent conference on Darwin and Evolution,(5), spoke of the achievement of an adaptation, say the eye, or even a red light sensitive cell in the progeny of an organism with no light sensitivity, as a 'blind teleology'.  Depew had in mind just what Darwin told us.  This is Monod's teleonomy - the appearance of design without a designer.  There is no doubt the eye is an adaptation - indeed the similarity of the vertebrate eye, so resembling a camera, is stunning.  As Monod forcefully points out, only life appears able to do this. Of course, the eye has evolved multiple time, but that is beside the point I raise. Other adaptations are unique. 

I now raise a central issue.  Can we speak of an opportunity for an adaptation before it occurs?  With thanks to G. Kaufman, (9), I translate such an opportunity for an adaptation, A, as 'A is possible. A might or might not occur. If A occurs it will \textit{tend to be selected} and fixed in the population'.   Now it becomes a critical issue to ask what kind of a 'cause' is the opportunity for an adaptation, which, if achieved, may change the course of biological evolution. It is clear that the actual achievement of the adaptation is via a series of efficient causes.  However, the \textit{tendency to be selected is a dispositional term}, untranslatable into any finite set of necessary and sufficient efficient cause conditions, or actual events, for the achievement of fixation of the adaptation. This means that we cannot state ahead of time the efficient causes by which a particular adaptation will come to be achieved.  But is the opportunity for the adaptation itself, \textit{the very fact that the eye is an adaptation} - subsequently achieved by non-prestatable efficient causes -  itself an efficient cause?  Certainly the opportunity for an adaptation is not an efficient cause in the straight forward sense of billiard balls hitting billiard balls.  Nor in more sophisticated terms, such as Maxwell's equations which are descriptions of efficient causes, is the opportunity for an adaptation in any clear sense an efficient cause.  Further, for a system with a potential, such as a ball rolling down a warped hill, where a least action principle can be found, that least action gives a superficial appearance of a final cause.  But there is no hint that the achievement of an adaptation is a flow on a potential for which a least action principle might be found.  And again the 'tendency' to become fixed by selection in the population is not, as noted, reducible to any actual (efficient cause) events that are necessary and sufficient for fixation to occur. Thus it does not seem that  we can translate the opportunity for an adaptation into any set of e
fficient cause events.

Addy Pross has given a very interesting analysis of this issue in the biological realm, (10,11). He distinguishes between thermodynamic selection, which tends toward thermodynamic equilibrium, and 'kinetic selection' among replicators for those which maximize a kinetic stability, not a thermodynamic stability. Pross's central point is that cells, as open thermodynamic systems, are unstable thermodynamically, but kinetically stable - they are the winners of a kinetic race in a 'space of replicators'.  I think Pross's insight is important, for in the simple case of, say bare replicators, the opportunity for an adaptation is a means to replicate faster to higher copy number, hence, as he says, higher kinetic stability.  Yet Darwin's fully biotic selection, and the economic selection of goods and services which survive in the market place, both analogous to Pross's kinetic selection, may go beyond any simple sense of 'winning the kinetic race'.  A butterfly may forego more rapid 
 reproduction if 'K' selected for carrying capacity in a nutrient limited environment, rather than 'R'  selection for replication rate.

I comment that David Deutsch (12) has written extensively on quantum mechanics and evolution.  

I give next four examples, two economic, then two biological, also referred to below in the section on enabling constraints, to argue that opportunities for adaptation are blind final causes in the case of the biosphere, and full Aristotelean final causes in the case of the economy, with the assumption of responsible free willed economic actors in the latter case.  Consider the following economic facts. In the early 1980s in North America, there were many television stations, abundant programming, many television sets, and, perhaps sadly, a multitude of couch potatoes.  In the face of this economic niche, was there an opportunity to invent and successfully market the television remote channel changer?  Yes of course there was and one could obtain venture funding to do so.  Now I ask, was the economic niche mentioned above an \textit{efficient cause} of the invention of the television remote?  No, it was, rather, as described below, an enabling condition, or enabling constraint, that a
 fforded the opportunity to invent and make money with the television remote.  Now consider the following: In 1943, the computer was invented to calculate shell trajectories in World War II.  Some thirty years later, the invention of the computer afforded the opportunity to invent and market widely the personal computer. IBM and Apple made substantial money on the venture. With the invention of the personal computer and its wide sale, the opportunity arose to invent and market word processing, and Microsoft made money doing so.  But the invention of word processing afforded the opportunity to store word files. In turn, stored files afforded the opportunity to share files between CERN colleagues, which in turn led to the economic-technological niche opportunity to invent and spread the world wide web. In turn, the web afforded the opportunity, the niche, where web commerce could find a home and EBay flourished.  In turn, the abundance of information on the web created the oppo
 rtunity, the economic niche, for the invention of web portals such as Google.  Now we have achieved the summit of Western civilization with Facebook.  Or, consider the flourishing of "aps" on cell phones and the growth of text messaging. Note how each of these opportunities, or enabling conditions, created a niche into which the next invention made economic sense. But the opportunities were not efficient causes of the inventions.

It might be thought that the story above relies on human conscious invention, but the same processes obtain for the evolution of the biosphere.  Organisms occupy niches.  As new organisms evolve, new niches are created. But a niche, for example, that occupied by rabbits, is not an efficient cause of the evolution of rabbits to fill and persist by existence in that niche.  Rather, the niche is an opportunity which evolution blindly seizes, and adaptations to fill that niche arise and are selected by efficient cause events as the adaptations tend to be selected by natural selection. The niche is not an efficient cause of those adaptations, although the actual steps of adaptation are themselves achieved by efficient causes.  Rather the niche is, as emphasized below, an enabling constraint that allows rabbits to arise and 'make a living' in that niche. 

A wonderful further set of examples arise in co-evolution.  Consider flowers, insects and birds such as humming birds.  Flowers feed the birds and insects nectar. Pollen rubs off on the insects and birds, is transferred to another flower and pollinates the latter.  Each is the niche of the other, and flowers, insects and birds have co-evolved their mutual niches for millions of years.  Step by step flowers found new adaptations to attract insects and birds and manage to be fertilized by insects and birds, and the latter adapted the stickiness of their hairs and beaks for pollen, and food gathering behavior, to carry out that fertilization.  The adaptation steps were achieved by efficient causes. The wondrous mutual emergence of the diversity of flowers and insects and humming birds as mutual co-evolutionary adaptations of ever creating niches is not  efficient cause.  Each of the mutualists gradually builds new opportunities for the other in their evolutionary becoming.  The Buddhists would call this 'co-dependent origination'. 

Physicists seeking a theory of everything from which all is entailed by deduction cannot ignore the biosphere's becoming, let alone culture, economics, and history where we become confused about consciousness and free will. Yet the evolution of the biosphere, say before consciousness evolved, is squarely in the purported purview of the physicist such as Weinberg.  But he cannot deduce this becoming, for opportunities for adaptations are not efficient causes, yet, once achieved by efficient causes, alter the course of evolution of the biosphere.  

I conclude that the opportunity for an adaptation is an opportunity for natural selection to select what will succeed in the current selective environment, and is a blind final cause, not an efficient cause. It may be pointed out here that with Darwin and with ourselves, it is essential that we feel it appropriate to use the phrase 'succeed in the current selective environment'. The red spotted organism will be a winner in Darwin's struggle for existence. But the very phrase 'struggle for existence' is, as philosopher Dan Cloud pointed out to me, to place the process of natural selection in a problem solving framework.  But a problem solving framework is not a mere description of what happens, as is the description of a ball rolling down a hill.  It is, in fact, true that the red spotted organism that is light sensitive is actually fitter than its non-light sensitive rivals in its selective environment.  This fact that this organism is actually fitter in the given environment
  is why - not how, but why - this fitter organism is selected. 'How' the selection actually occurs is a sequence of efficient causes such that the fitter organism dispositionally 'tends' to win. But we cannot state what those efficient causes must be. Again, I conclude that the opportunity for an adaptation is a blind final cause, not an efficient cause of what merely happens.  This is an essential step, for it claims that the becoming of the biosphere is not sufficiently describable only by efficient causes.  But this will imply that the becoming of the universe including the biosphere is not describable only by entailment from mathematicized efficient cause laws.  In turn, this means that we are not limited to the tautological entailments of a final theory of everything, and that an open creativity beyond entailment is present in the unfolding of the biosphere, economy, history, and perhaps the universe as a whole. 

I remark preliminarily that to speak of an opportunity for an adaptation, we seem forced to deal with the fact that the adaptation is 'possible'.  Already in Quantum Mechanics and the Schrodinger equation with Copenhagen and the Born rule, we speak of the Schrodinger wave as a 'possibility wave' which, when its modulus is squared, gives the probability of observing possibilities that we know beforehand. We will soon see that the evolution of the biosphere seems to force us to a wider possible, where we do not know beforehand what the possibilities are.  All this is puzzling. In General Relativity and the block spacetime universe, there are only world lines, actuals, and no possibles. We shall have to begin to inquire about the status of the possible.

\section{The Evolution of the Biosphere by Darwinian 'Preadaptations' is Partially Lawless}
 
Were we to ask Darwin the function of the human heart, he would say it is to pump blood. Were we to point out that the heart makes heart sounds and moves water in the pericardial sac, he would say these effects are not the function of the heart.  If we asked why not, he would reply that the heart was selected, so exists in the universe, because it was of selective advantage to pump blood in some ancestor and the lineage leading to us.

Already this is interesting because, were the physicist to succeed in deducing all the causal properties of the heart from its subatomic constituents, she would have no way to pick out pumping blood as the biological function of the heart and the putative reason hearts came to exist in the universe.  To describe the function of the heart, she would have to become a paleontologist and evolutionary biologist, or to simulate the evolution of the biosphere, or deduce from her theory of everything the emergence of the heart.  In two books, Investigations and Reinventing the Sacred,(13,14), I argue that she cannot simulate or deduce the emergence of the heart in the unfolding of the universe.  Quantum events matter in evolution, at least by causing mutations.  There is no way to simulate all the quantum processes that \textit{have occurred}, including random cosmic rays, or, in accord with Schrodinger,\textit{ might have occurred}, in the history of the past 5 billion years of the earth, let alone u
 niverse. How would one simulate the all the possible consequences of all the possible temporal instants of a radioactive decay, or a quantum coherent electron transfer in some protein in some organism in some environment? Now consider doing so for all the quantum events in the past 5 billion year history of the Earth and evolving biosphere. More there is no way to confirm that any such simulation captures the actual quantum history of this biosphere's evolution.  But can the physicist deduce the becoming of the human heart in evolution, or evolution more generally.  I now argue that the answer is a resounding 'No'.  If I am correct, it appears to have major implications.

Darwin spoke of the fact that a feature of an organism, with some causal property of no selective significance in the current environment, might be of selective value in a different selective environment, so be selected. Typically a new function will arise in the biosphere.  These events are called either 'exaptations' or Darwinian 'preadaptations'.  There is no concept of evolutionary foresight here. It just happens to turn out that a property that is of no selective use in one environment is of selective use in another environment.

I give two examples.  Some fish have swim bladders. These are sacs, partially filled with water, partially filled with air, that adjust neutral buoyancy in the water column.  Paleontologists believe that swim bladders evolved by exapatation from lung fish. Water got into the lungs of some lung fish, now there was a sac partially filled with air, partially with water, and so poised to evolve into a swim bladder.  Let us assume the paleontologists are correct.  Now: Did a new function arise in the biosphere? Of course, neutral buoyancy in the water column.  Did the swim bladder affect the further evolution of the biosphere? Of course, new species, proteins, other molecules, and niches evolved.
Here is a second example. We have three middle ear bones to transmit sound from our tympanic membrane to our inner ear. These evolved by preadaptation from three adjacent jaw bones of an early teleost fish.  This case is important because relational 'degrees of freedom' matter. If the three bones were not adjacent, but were in the spine, skull and jaw, probably middle ear bones would not have evolved. Again, did a new function come to exist in the biosphere? Yes, hearing. Did this new function alter the evolution of the biosphere?  Of course, new species, proteins, niches. 

I now come to the critical question:  Do you think you could prestate all the possible Darwinian exaptations of all organisms alive now?  You might respond that we do not know all organisms alive now.  I simplify my question: Do you think you could prestate all possible Darwinian exapatations just for humans?  I have now asked thousands of people. We all agree we cannot carry out this task.  Why not?  I think parts of the problem are that we cannot prestate all possible selective environments, nor know that we had listed them all. Nor can we prestate all features of one or many organisms, including relational features, that might turn out to be preadaptations.  It is not clear how to prove this claim. An experiment seems beside the point A theorem seems impossible at least at present.  

I now need to define the 'Adjacent Possible'. Consider a liter of buffer with 1000 different molecular species. Call this set the 'Actual'. Let them react by a single reaction step. If new species of molecules appear, call these 'The Adjacent Possible'.  Clearly this is well defined in the chemical case, given a minimal life time of stability for a species.  Now let me point to the Adjacent Possible of the biosphere. Once there were lung fish, swim bladders were in the Adjacent Possible of the biosphere. Before there were multicelled organisms, swim bladders were not in the Adjacent Possible of the biosphere.  Admittedly, I use some poorly defined sense of 'adjacent' here.

Now if we do not know all the possible preadaptations that might arise in the adjacent possible of the biosphere, then not only do we not know what \textit{will} happen, we do not even know what \textit{can} happen!  Can we make probability statements about the evolution of the biosphere by preadaptations?  Consider flipping a coin 10,000 times. It will come up heads about 5000 times, with a binomial distribution. But notice that we knew ahead of time \textit{all the possibilities}, all heads, all tails, and so forth. We knew the sample space of the process, so could erect a probability measure on the frequency interpretation of probabilities for this coin flipping process.  \textit{But we do not know the sample space of the evolution of the biosphere by preadaptations, so can make no probability statements about it.}  Now Laplace had a different interpretation of probability. If confronted by N doors, behind one of which was a treasure, but we had no idea which door, our chances of picking the right door is 1/N.  But notice that we know N, the number of doors. We do not know N for the evolution of the biosphere, so can make no probability statements about this process.

If a natural law is a compact description of the regularities of a process, can we have a sufficient natural law for the emergence of swim bladders?  No.  We cannot even state the possibility of the emergence of swim bladders, let alone their probability. Thus we cannot have a law that is sufficient for describing the emergence of swim bladders.  

This is a major conclusion. \textit{The becoming of the biosphere is partially beyond sufficient natural law.  Yet it is also non-random. } There is no sufficient law for the becoming of the swim bladder, yet this new organ does make sense and is selected in its selective environment, hence its evolutionary emergence is not random.  We have no such concepts in physics of a partially lawless yet non-random process. But the biosphere appears to be doing just this.  The same is true in the economy, culture and history.  
But if the emergence of the swim bladder is not describable by sufficient natural law, it is not entailed by any theory of everything at the fundamental level of physics.  Thus, there can be no theory of everything!  Nor can the evolution of the biosphere be deduced by mathematized efficient causal law.  This failure reinforces the conclusion that adaptations are blind final causes, and our explanations of the becoming of the universe are not limited to efficient cause laws.  Notice that this discussion is not that of Hawking about Godel and the End of Physics based on efficient cause mathematical law and the possible inadequacy of any finite set of such laws, which may also be valid in its own right.

It is important to pause for a claim about 'the furniture of the universe'.  Are swim bladders ontologically 'real'?  Consider proteins length 200 amino acids. How many are possible with 20 kinds of amino acids? 20 raised to the 200th power, or about 10 to the 260th power.  We can make any one of these we choose. But were the 10 to the 80th particles in the known universe to do nothing, ignoring space-like separation, on the Planck time scale of 10 to the -43rd seconds but make proteins length 200 amino acids, it would require 10 to the 39th power repetitions of the history of the universe to make all these proteins just once.  But this means that, at levels above stable atoms, the universe is on a unique, utterly non-ergodic trajectory.  Most complex things will never exist, so the existence of the heart is no small matter.  But if we cannot deduce the coming into existence of hearts or swim bladders, and yet they have causal powers as organized structures and processes, then hearts and swim bladders are emergent with respect to the fundamental laws of physics and so are ontologically real parts of the universe.  We are not just particles in motion. Moreover, since most complex things will never exist, the universe is indefinitely open upward in complexity. And since efficient causes, mathematized as deductions, do not suffice to describe the unfolding of the universe including the biosphere, the universe is open and, for the biosphere and upward, vastly creative.

I also pause to note that the richly interwoven complexity of the biosphere which has emerged cannot be captured by Shannon information. Shannon assumes an ensemble of messages, in a prestated alphabet, where all possible messages are known beforehand, and thus whose entropy can be calculated. But we do not know the all the possibilities that evolution will unfold. We do not know the alphabet of processes, entities, and functions that will emerge and integrate into an evolving biosphere. Whatever information may be, a vexed question, Shannon information does not seem to apply to the evolution of the biosphere.  Indeed, I do not think that this evolution is even algorithmic, (13,14).  Consider the famous Halting problem, where no compact description of the behavior of some algorithm may be available. But for the next 11 steps, or any finite number step of the universal Turing machine, all possible states, in a prestated alphabet, of tape and head, can be listed.  We cannot even get started on the evolution of the biosphere by preadaptations.  So our problem with the evolution of the biosphere does not seem to be the same as the problem of there being no compact description for an arbitrary algorithms behavior. We may well confront the issue that no language describes all of reality.

I end this section with an economic preadaptation, said to be a true story. Engineers were trying to invent the tractor, so knew that a massive engine block would be necessary. This was placed on a succession of chasses, all of which broke. At last an engineer said, "You know, the engine block itself is so big and rigid, we can hang everything off the engine block and \textit{use it} as the chasse".  This novel use of the engine block is a Darwinian economic preadaptations.  Economic inventions are rife with similar examples and most inventions are not used for their initial inventive purpose.  This raises the issue of algorithmicity again.  Can we name all uses of a screw driver? No. This is the 'frame problem' of computer science, never solved.  I think that the human mind, like the evolution of the biosphere, is not algorithmic, and the evolution of the economy, culture, and history are not describable by natural laws, (14). Indeed, historians, who do find out about the real world,  today largely eschew a search for the laws that Marx sought.  In part, history and cultural evolution, like the invention of Google, are instances of opportunities seized - not merely efficient caused events. 

\section{Is the Coevolution of the Quantum Classical Boundary Lawful?}

As we consider the adequacy of reductionism, it becomes of interest to ask if the boundary between the quantum and classical worlds, their co-evolution, is lawful.  In this section I borrow an argument from Sir Karl Popper in his The Open Universe, (15), to suggest that this becoming is not describable by sufficient efficient cause law. The ideas have testable consequences, in principle. 

In Popper's argument, the setting is Special Relativity. An event A has a past light cone and a future light cone, with a zone of possible simultaneity between them. An event B is in the future light cone of A. The past light cone of B includes the past light cone of A, but includes regions that are space-like separated so lie outside the past light cone of A.  Popper then argues that, at A, we cannot know the events in the past light cone of B that are outside the past light cone of A but may influence event B, so we cannot have a law for the event B before B occurs.   \textit{If a law is a compact description of the regularities of a process which an observer at A, and before event B,  can construct,} Popper's argument seems valid.  If the observer is not located at A, then we will be driven to an observer outside the universe, which seems inadmissible. Popper uses his argument to support indeterminism.

My own setting depends upon the currently popular theory that the transition from quantum to classical is due to decoherence and loss of phase information from the system to an environment, quantum, classical or both. The loss of phase information to the environment means that the system gradually loses the capacity to exhibit interference patterns like the two slit experiment, the hallmark of quantum behavior. The transition to classical behavior is often described as 'for all practical purposes', (FAPP), since the system's phase information continues to exist in the environment. Take a setting like Popper's. For example consider a complex organic molecule in a dense solution of such molecules, and an event A in which two emitted entangled quantum degrees of freedom that move apart from that molecule and eventually are absorbed by one or two detectors,  event B, say classical, that recede from one another at constant velocity, the Special Relativity setting.  Then Popper's argument applies.  \textit{Before} the absorption by classical or other quantum degrees of freedom, we cannot know what events outside the past light cone of the complex molecule, event A,  and the receding quantum entangled particles, may impinge upon decoherence upon absorption, event B, and EPR instantaneous correlation with the quantum decohering molecular system from event A.  Then we do not know how decoherence happens in detail in that molecule.  Then there can be no efficient cause function, F, or law, for the detailed way decoherence of parts or all of the complex molecule happens. Thus, it appears, we can have no law for detailed decoherence in this Special Relativity setting.  But quantum mechanics and Special Relativity are consistent, as Dirac's relativistic electron equation argues. This claim implies that there is no efficient cause law, or function, that maps the space time region including event A and the receding detectors before event B, into a future that includes 
 event B. 

If there is no law, what can we say about what happens?  I discuss this below. If there can be no law, then it seems there can be no Theory of Everything from which all that happens is entailed. 

There can, of course, be statistical \textit{models} of this decoherence process. But such models are not detailed laws. However, if the above view is correct, it seems to vitiate full reductionism - the dream that there is an efficient cause law or set of laws that entails all that happens in the universe.

The situation is even more complex, for the transition from quantum to classical (for all practical purposes if you wish) and back is thought to be reversible.  Shor's code for error correction in quantum computers, (16), shows that in a quantum computer, decohering degrees of freedom can be made to recohere with addition of information from the outside.  H. Briegel has recently published two papers, (17,18), arguing that a quantum entangled system can become classical then fully quantum entangled again.  Assume Shor and Briegel are correct.  If decoherence is lawless, then even if the classical to quantum transition is lawful, the total quantum to classical to quantum reversible process must be lawless.  \textit{But that means that the coevolution of the quantum-classical world is not describable by efficient cause mathematical laws.  Again, it seems there can be no Theory of Everything.}

Given our interest in Darwin and natural selection, it becomes of considerable interest that a speculative \textit{abiotic natural selection process may arise at the quantum-classical boundary.  Decoherence seems likely to depend upon the local quantum plus classical environment. The more complex the environment, presumably the easier and more rapid decoherence of the system will be.  Then quantum degrees of freedom that have decohered to classicity for all practical purposes, and are more resistant, in that complex 'selective environment'  to returning to the purely quantum condition, will tend to persist as classical entities in the universe. This will depend upon the local 'classicity selective environment' and is a possible form of abiotic natural selection with abiotic blind final cause due to the (possibly changing)  selective environment.} This argument supplies the start of an answer to 'what happens' if there is no efficient cause function, or law, at the quantum-classical boundary.  For, as in the case of biological evolution, the selective environment determines in part how readily a now classical entity tends to remain classical rather than becoming quantum again by recohering. 'Tends' is again a dispositional term. The actual ways that decoherence happens and is sustained against recoherence will depend upon actual detailed quantum and classical processes.  We cannot prestate those selective environmental processes that are necessary and sufficient for the now classical (for all practical purposes) entity to remain classical, FAPP.  Thus, there is a process carrying the system into the future, but no efficient cause law, or function, describing it.  

The above should be experimentally testable.  In general, it is now believed that complex entities decohere more rapidly than simple entities, eg electrons and photons, which means a bias towards the emergence of classicity in complex entities. Abiotic natural selection arises here with blind final cause, for we cannot prestate all the complex environments which may impact decoherence in any specific way.  Anton Zeilinger has recently shown that Buckmeisterfullerenes interfere in a two slit-like experiment. Presumably, \textit{as the complexity of the objects in this experiment increases, and the complexity of the surrounding environment increases, decoherence should begin to fail.  As it does, it may be possible to ask whether the decoherence process is fully lawful or not, for example, by failure of stable statistics in fading interference bands. More, if the complexity of the environment bears on decoherence, then at that molecular complexity when interference begins to fail due to decoherence, one would expect that a dense 'beam' of the objects sent through the two slits would behave more classically and show less interference,  than if the objects were sent through the two slits rarely.} It seems that the above are possible new experiments. 

Finally, I note that D. d'Lambert commented to me that the above ideas imply that the quantum measurement problem does not have a solution, (19).  Taken together these ideas, if correct, again seem to imply that there is no Theory of Everything from which all is logically entailed. I comment that W. Zureck might strongly disagree, (20).

\section{Enabling Constraints and What They Enable}

In about the year 1200 AD, the Calif of Cairo caused the only hospital in the Islamic world to be constructed.  Because patients were required to be treated within the hospital, where Maimonides later practiced, it became possible to train medieval physicians in a new manner.  The hospital enabled a new form of medical education and medical practice.  More, the Calif was able, as a sign of caring, to visit the patients in the hospital and thereby talk to poor people he could not have met socially.  This allowed the Calif to gain different information about his realm and govern differently.

The hospital acted as an enabling constraint or enabling condition, and enabled changes in medical education, treatment and governance.  We obviously know this is true, but have virtually no clear ways to think about enabling constraints or what they enable.

A second example was raised by A. Juarraro in Dynamics in Action, (21). Could we cash a check 50,000 years ago? No. Think of all the social inventions that had to occur to allow this bit of human action.  Laws, courts, credit, bankruptcy laws, enforcement procedures, contract law, all had to come into existence.  In the law, the concept of enabling constraints is known.  If you and I enter into a contract, we are thereby constrained, but may be enabled to form a corporation by that contract, with all the enabled actions of a corporation in the contemporary world.  The stories above of the invention of the television remote, and the sequence leading from the first computer to FaceBook, are also examples of situations arising that create new economic niches and are enabling constraints, but not efficient causes.  The enabling constraints create opportunities seized. We know this is true, but do not think about it. We have no theory for it.

Enabling constraints arise in biological evolution. A signal case is the evolution of meiosis, chromosomal recombination and sex.  Sex causes a two fold loss in fitness as two parents are required, not one.  But sex allows meiosis and chromosomal recombination between homologous paternal and maternal chromosomes that permits two advantageous genes, say A and B, initially with one on the maternal and one on the paternal chromosome, to recombine so A and B are on one chromosome and passed via sperm or egg to the offspring.  This process is much faster than waiting for A to arise by mutation on the B bearing chromosome, so abets more rapid and efficient evolution.  In short, sex is an enabling constraint!  The biosphere is not only evolving, it is evolving the way it is building itself. It is evolving the very way it is evolving. If sex and recombination yielded the emergence of Mendel's laws, then life evolved its own enabling constraint laws by which evolution itself became more efficient.  Then might the universe as a whole evolve its laws so that its becoming' was more efficient in a form of a Darwinian race among a set of candidate enabling constraint laws and some definable notion of 'efficient'?   

A second biological example almost certainly arose early in life.  Current cells use DNA, RNA, and encoded protein translation. But the process is very complex, with transfer RNA and specific protein enzymes each of which charges the appropriate transfer RNA with the 'right' amino acid to allow proper translation of messenger RNA.  The entire system is needed for the system to work.  Early in the evolution of life, proto-cells presumably were reproducing, perhaps did work cycles, but could not have been so complex.  Call the emergence of DNA, RNA, and encoded protein synthesis 'the Darwinian Transition'. This transition has become an enabling constraint. All of life since, presumably, the last common ancestor, has used this molecular machinery: we are constrained to it. Yet this machinery enables the rapid exploration of protein space by mutations to DNA sequences not needed for core molecular reproduction.  The biosphere, again, is evolving the very way it evolves.  The DNA/RNA/protein translation machinery is a powerful enabling constraint 'law', the central Dogma of molecular biology,  that has enabled enhanced evolution.

These biological examples seem deeply important for, unlike human law, no conscious agency is invoked. The biosphere is building the way it builds itself by evolving law-like enabling constraints that enable enhanced biological evolution.  This is an existence proof that nature is able to achieve such a miracle. We broach the universe as a whole below.

I do not believe that these evolutionary processes are algorithmic, (13, 14).  We have no theory of enabling constraints and what they enable, which I also do not think are algorithmic in general. I have no idea how to study enabling constraints and what they enable in general, so I now sketch the earliest stages of an admittedly limited and algorithmic approach to this question that is now underway.

Consider chess.  The rules of chess are the enabling constraints, the laws of the world of chess.  They enable very sophisticated, strategic play, as many of us more or less know.  We do not understand, in general, enabling constraints, and what poor or superb 'strategies' can emerge as in the history of chess play.  But a few observations start a discussion. Note that, given the move rules of chess, the Adjacent Possible for White, or for Black, is fully determined for each board position.  Then I propose to ask, as a game proceeds, what happens to the 'size' of the adjacent possible for each side.  In the end game, typically the losing side has almost no adjacent possible, while the winning side has a very large adjacent possible. How does this happen?  How is it related to the search depth of computer chess programs playing one another, whether of equal 'strength' or different strengths?  I intend to find out in this simple case.

Note that in chess the bishop can move along a diagonal that is free, regardless of the position of the same side's rook, as long as the rook does not block the diagonal. The movements of chess pieces are largely independent of one another except for blocking.  But one can imagine chess - like rules in which all positions of the rooks impacted the legal moves of the bishop.  Or in which all positions of all pieces impacted the allowed moves of the bishop. As one tinkers with the move rules, and the dependencies of pieces moves on one another's positions, what happens to the games that are enabled?  What happens to the adjacent possible?  We don't know. Are the most complex games achieved, under a to be determined criterion of 'complex', if the pieces moves are largely independent of one another?  I have no idea.

My colleagues and I are also starting work on board games in which each side has M pieces, each piece has, for each board condition a set of allowed next positions.  Thus, the adjacent possible for all board positions is perfectly defined. As we tune the dependency of each piece's moves on the positions of is own sides other pieces, what happens to the adjacent possible of each piece and why?  If we start in the same position and step randomly several steps into the successive adjacent possibles of a piece, and repeat this sequence many times, do these 'histories' spread out widely? Do they converge?  Are there some board positions reachable by very many other board positions, and others that are hardly accessible?  If so why?  What are the implications of this possible variation in adjacent possible board positions on the flow of the games we envision next.  

We plan to study games where each side can 'take' a piece from the other side by occupying its position via a legal move, as in chess or checkers.  We propose to allow two depth search, so each side can both 'try' to take an opponent's piece and try to avoid having its own pieces taken. A game will be won when all pieces of one side are taken. We propose to evolve the rules of moving the pieces, so that winning players (or both winners and losers) can alter their move rules to a set of 'next move rules' to evolve toward rules that allow longer more complex games. One measure of the complexity of a game is to replay the same game multiple times, treat a board position as a vector, and concatenate successive board positions of one game until the game is won into a long vector. Repeated games will give some diversity of these vectors. The 'normalized compression distance', (22), between many pairs of games can then be computed, to gain a measure of how diverse games under a given set of move rules are.  This diversity is one measure of game complexity.  As the move rules evolve toward more complex games, we hope to look at the dependency of each piece's adjacent possible move space on the positions of other pieces of the same side.  I hope we find that 'complex games' evolve relative independence of one piece's moves on another's positions except blocking positions.  

In short, a new body of theory is needed where virtually none exists: What are enabling constraints and what possibilities do they enable?  Board games are interesting because they are so well defined.  They are inadequate because the move rules enabled by the Cairo hospital were not algorithmic, as are the board games. An entire new field of research is needed. I believe and feel sure it is worth exploring. All our legal codes, regulations, the biosphere and perhaps, as I try to discuss next, the very physical laws of the universe, are enabling constraints. What do they enable? How?  

Notice for further discussion, that the move rules define an Adjacent Possible. Below I ask where does 'the possible' come from?

\section{Might the Laws of Physics Be Abiotically \\ Selected Enabling Constraints?}
 
Where are we now in fundamental physics and cosmology?  We have the Standard Model and General Relativity, and as yet no clear way to unite them.  If the above argument about lawlessness and abiotic blind final cause at the quantum classical boundary is right, we may never unite the two. If blind final cause is present in the evolution of the universe including the biosphere, let alone human culture, there may be no theory of everything entailing all that occurs. Meanwhile, we have the well known 'fine tuning' of the 23 constants of nature.  It is widely believed that without this fine tuning we would not be in a complex universe with stars, simple and complex atoms, chemistry and life.  But we have no rationale for why the constants have the values they do.  

In face of this fine tuning, the current view in physics is of a multiverse, where each 'pocket' universe has its own values of the constants, perhaps randomly distributed, and either the strong or weak Anthropic principle.  The former looks to a Creator God to tune the constants and is held to be outside science. The latter assumes that only those universes with constants disposed to allow stars, complex atoms, chemistry and life would have physicists to puzzle about why the constants of their pocket universe were so tuned as to allow their existence.  Probably the weak Anthropic principle is the dominant view among physicists today.  Leonard Susskind, confronted with 10 to the 500th string theories, envisions a cosmic landscape, with as many pocket universes, each with a random choice of string theory from among the 10 to the 500th, and we are the lucky ones, (23).  Lee Smolin, in Life of the Cosmos, (24), imagines universes born from black holes and emerging with minor variations of the constants, so a cosmic natural selection among universes for those that are more fecund because they have many black holes.

It is worth stressing that reductionism itself is what is driving us to the multiverse.  If we cannot account for the fine tuning of the 23 constants of nature, and if all that arises in any universe is deductively entailed in its efficient cause Final Theory of Everything, there is no choice but some space of possible laws or one set of laws but many choices of values of the 23 constants, multiple universes, and some way of distributing the laws, or constants, among these universes.  But, as I note in a moment, Darwin tells us that we are not limited to efficient causes, and that may change everything.

I now propose 'Darwin all the way down'. Suppose that there was, in the beginning, or in a 'possible' before the beginning as I try to discuss below, an indefinitely or infinitely large set of laws to create universes - I'll give a conceivable example in a moment - and a cosmic natural selection selected, \textit{in just one universe}, those laws which, as enabling constraints, enabled our very complex universe precisely because it was able to grow large and complex, hence by persistent winning 'existence', won  Existence and persistence are the abiotic analogues of the persistence of saber tooth tigers existence and persistence in the biosphere. Existence and persistence of a 'winning' universe that does so 'the best', is the analogue of Pross's kinetic selection in a non-equilibrium chemical replicator system. We will see and are the winners. 

We then are in this universe, because, Darwin-like, it is the universe that won by blind final cause.  We now answer, in principle, a why question and answer not just with an efficient cause 'how' answer, but a 'why' answer. Our universe won and was able to become a very or most complex universe.  That is why our universe is as it is.  For us to be satisfied, what constitutes 'winning' for a universe must itself be 'natural'. For example, I want to believe that the biosphere evolves, as a secular trend, to maximize its Adjacent Possible in the non-ergodic universe: Perhaps as species diversity and features per species and complexity of features increase, the ease of forming positive sum games and mutualisms increases, driving further diversification of organized processes in the biosphere.  Perhaps the winning universe wins by maximizing its Adjacent Possible into which it can 'become' more rapidly than universes that grow their Adjacent Possibles more slowly.   Like the biosphere, the universe as a whole is vastly non-ergodic. The metaphor is at least suggestive. Science sometimes starts with a mere metaphoric image that later crystallizes usefully.  The metaphor of the solar system for atoms is a famous example.

\textit{We must note that any effort along these lines is radically unlike familiar physics, for we are attempting to formulate the question: how are physical laws enabling constraints and what kinds of universes do they enable?  And if there are a multitude of laws, how might an abiotic natural selection process with blind final cause work to select among the laws? And what constitutes a "winning" universe that might, by blind final cause, select the laws that enable it?}  

I briefly mention a 'vacuum selection' principle of which I am the author, (13).  It is only of interest as an example to show that such a vacuum selection principle might be possible.  Smolin and colleagues have explored loop quantum gravity, (25). Here Planck scale tetrahedra of quantized units of space build a universe by budding or cloning new tetrahedra on their faces, via Pachner moves, where the tetrahedra are linked by what are called 15J symbols. As Louis Crane showed, these 15J symbols, all integers, form a denumerably infinite series of laws, (26), so can be pictured as a space of laws with an ordering relation among them. Each 15J symbol implies the way the discrete analogue of the Schrodinger equation propagates on the space constructed by the tetrahedra.  My idea was to allow \textit{uncertainty of the laws themselves} in an early universe, with a universe starting in one state of geometry, (and ultimately particles), with one 15J symbol, and following all possible paths
  to a final state where the 15J laws were different.  Thus, \textit{if the particles under these different laws, or geometries themselves, could interact, quantum  interference could arise.} I reasoned that \textit{some small changes in the 15J symbols could yield large changes in Schrodinger propagation, hence yield destructive interference. Other small changes could yield, I hoped, very small changes in how the Schrodinger equation propagated possibility amplitudes, so lead to constructive interference.}  More slowly, Feynmann showed in his sum over all possible histories formulation of quantum electrodynamics, that nearly parallel pathways interfered constructively, while radically twisting pairs of pathways interfered destructively, so near classical parallel behavior was the most probable.  Generalizing to the case where there is to be uncertainty over the laws themselves, and summing over all histories from all initial to all final states of tetrahedral space, with the same or \textit{different} 15J symbols, I hoped, mere sum over all possible pathways and constructive interference, as Feynmann showed with a single Schrodinger equation, would pick out the region(s) in the denumerably infinite space of laws where constructive interference among a neighboring set of laws would arise where small changes in 15 J symbols yielded tiny changes in how the Shrodinger equation propagated, hence a universe whose laws, if initially fluctuating slightly, showed constructive interference, would arise.  Ultimately, this universe would select out a single law 15J law. This simple example, merely conceptual, is the start of a possible vacuum selection principle among an infinite set of laws in a single universe with an infinite set of possible laws, yet might be able to select the laws and ultimately, I hoped, the constants, the particles, and all. If even logically possible, this putative vacuum selection principle suffices as an example of a way a single universe might evolve its own laws. No multiverse is needed here.  If not, we are not forced, e
 ven by reductionism in the sum over histories and laws above, to a unique or very small set of neighboring laws, to posit a multiverse.

But there may be other vacuum selection principles evolving the laws if we allow forms of blind final cause for a 'winning universe' selecting among the possible laws, all in one universe evolving its laws so it 'becomes' better.  In the case above of my hoped for vacuum selection principle by constructive interference over sets of 15J laws, we already have a Feynmann framework to understand what 'winning' might mean - constructive interference.  As noted, we need to explore a wider set of what a winning universe that exists and persists might mean and how that could "blindly" select among enabling constraint laws that enable its Adjacent Possible.

Since we do not know what enabling constraints enable what kind of universe, only hints are available now. What might they be? It would seem, as noted just above, that 'getting to persist' - like saber tooth tigers -  would be important.  Perhaps relative local independence of classical events, like the bishop's moves independent of the rook or the same in complex board games, might be essential to a winning universe that can become big and complex, or maximize the growth of its Adjacent Possible, hence win.  There are  clues to such 'move independence'.  Nother's theorem, (27), shows that where there are symmetries, for example of force applied and acceleration achieved, with temporal, translational and rotational invariances, conservation of energy, momentum and angular momentum are entailed.  Why should these independencies with respect to these spatial and motion symmetries be a feature of our laws of physics?  Bishops and rooks? Does this enable a universe with a larger Adjacent Possible? Perhaps, in due course, this still intuitive question can be formulated precisely.

Our particles form a group, with its symmetries. \textit{The group property implies that the particles transform into one another, hence persist.} What if particles did not do this, but transformed into a spray of ever new particles such that, even were reversibility allowed, they created an infinite 'jet' of particle types that would emerge. Then nothing would persist.

Could such group particle properties emerge from the evolution of random laws?  No one knows. But I now report remarkable results that hint the answer could be yes.

I describe a wonderful numerical experiment some years ago by Walter Fontana, (28), at the Santa Fe Institute. Fontana created a 'chemostat' on his computer which contained up to 50,000 Lisp expressions. Lisp expressions were chosen randomly to act on Lisp expressions typically yielding new Lisp expressions. Selective conditions were maintained by randomly throwing out Lisp expressions if there were more than 50,000 in the computer chemostat.  Fontana found that at first a stream of unique Lisp expressions were generated. Then one of two things happened. First, a Lisp expression able to copy itself emerged and took over the chemostat. If copying was disallowed, \textit{collectively autocatalytic sets of Lisp expressions emerged, in which each was formed by one or more of the Lisp expression present.}  Fontana found that these collectively autocatalytic sets of Lisp expressions \textit{formed an algebra}, but not a group in that they lacked an inverse and an identity operator. Nevertheless, his  numerical experiment is a toy example of entities bootstrapping themselves via random laws co-evolving into self consistent co-creation and stable existence and persistence.  It is a long way to elementary particles forming a group, and transmitting forces, but perhaps a hint.  The transformations among the Lisp expressions are mediated by Lisp expressions and seem the analogue of forces carried by particles acting to transform particles into one another in a group. If models can be explored which include the possibilities of reversible transformations mediated by the same 'expression' acting on two interconverting pairs of 'expressions', and a 'do nothing' expression, perhaps autocatalytic sets of expressions might emerge from a soup of co-evolving random 'laws' or 'expressions' and form algebraic groups, (29).

Why are there conservation laws like that for a perfect harmonic oscillator, free of friction?  In the state space of position and velocity, orbits are concentric curves.  Adjacent curves have 0  Lyapunov exponents.  Thus, they are dynamically \textit{critical}, and can persist and propagate information without loss due to convergence in state space nor, in a noisy world, loss due to a positive Lyapunov exponent and chaos. Why?  Electromagnetic waves propagate, exist and persist therefore, across the universe. They can propagate information extremely well. Why such conservation laws?  What do they enable? Do they enable a larger Adjacent Possible for an emerging universe? 

We do live in an extremely complex universe.  If we take the fine tuning arguments seriously, this is a profound puzzle.  Just perhaps abiotic natural selection provides a radical but ultimately useful new way to think about this.  If so, Godel is not the end of physics, but all this is a possible new beginning in an open universe.

\section{The Possible}

How can we begin to think about 'the possible'? It seems we have to consider the possible and its ontological status. We seem to need 'a possible'.  I will proceed in steps based on physical theory and beyond.

First, consider General Relativity and Einstein's block universe. Here there are no possibilities at all, only actual geometric world lines.  Next consider Newton, where a state of the system in space and time has a possible future and past deterministic trajectory.  It is not much of a possible, but more than in General Relativity. Next consider quantum mechanics on the Copenhagen interpretation and Born rule. Here we have at the fundamental level, a Schrodinger equation for possibility waves. So we seem forced, on Copenhagen at least, to consider the 'possible'.  But notice an odd fact. We know beforehand exactly what the quantum degrees of freedom are, spin, polarization, and so forth, that we will measure.  Whitehead, in Process and Reality, (30), considers a metaphysics of Actuals giving rise to Possibles that give rise to Actuals. But in quantum mechanics, such as Quantum Electrodynamics, possibles can give rise to possibles in Feynmann's sum over all possible histories
  and his famous Feynmann diagrams.   

Now consider the evolution of the biosphere by Darwinian preadaptations. We seem to confront an Adjacent Possible of the biosphere where, unlike the possibles of quantum mechanics, we cannot prestate the relevant degrees of freedom, eg swim bladders.  Unlike familiar quantum mechanics, we do not even know what the variables might be.  This failure may be due to the failure of human language to describe all of 'relational' reality in a continuous spacetime in a denumerably infinite language. In any case, we seem forced to consider 'the possible', even one we cannot prestate. The same is true for the evolution from the computer to FaceBook and history with its Cairo hospital. Who foresaw the changes in medical training and practice that were enabled, became possible, then actual?

'The Possible' produces confusion because, in part, we live among Whitehead's Actuals. As with consciousness itself, we don't know 'where' the possible is in space and time. 

Consider any physical theory which posits a multiverse with a set of possible values of the 23 constants, or Susskind's Cosmic Landscape with pocket universes having one of the 10 to the 500th string theories.  Then it seems we are forced to consider a set of 'possible' values of the constants, or string theories, somehow assigned to, or coming into existence with, universes in the multiverse. What sense does it make to speak of these possibles before there are any universes? It seems physicists may have slipped into speaking of these possibles as, in some sense, real possibilities,  outside of any universe(s) that exist, whatever that means. Can there 'be' a 'possible' before' one or many universes exist and out of which it or they can become?  Can a 'possible' make any sense without the enabling constraints that seem to define it?

If we can speak of possible values of the 23 constants assigned somehow to pocket universes, or 10 to the 500th string theories assigned somehow to pocket universes, then it seems no stranger to consider a space of possible laws, for example the 10 to the 500th string theories, before there is our one universe and a very rapid vacuum selection principle, perhaps like mine above, that, in a single universe, selects by constructive interference among competing laws, or by blind final cause, that universe that 'wins'.  Like the weak Anthropic principle, such a vacuum selection would answer the question why the constants have the values they do. And it would answer the question: why these laws and partial lawlessness. But as noted, to base our thinking on an abiotic natural selection among a family of laws, perhaps infinite, to answer this question means understanding what a 'winning' universe might be, how the enabling constraint laws enable that winning universe, and how it wins over other universes in the early evolution of our one universe and thereby selects its own laws. Like the evolution of sex and reconbination and Mendel's laws, are our physical laws 'enabling constraints' that became hardened and entrenched as the universe itself evolved, such that the universe was then constrained to those law? If the discussion above, in whole, is correct, to pursue this avenue means giving up the reductionist dream of a final theory, but it may open wide new doors in an open creative universe.

\section*{Conclusion}

Reductionism has been a brilliant success. It is built upon the mathmatization of efficient cause and that cause as deductive entailment.  It appears that this is insufficient to describe the becoming of the biosphere by adaptive evolution and more the evolution of the biosphere by Darwinian preadaptations.  The considerations above, together with Hawkings Godel and the End of Physics, suggest we may be approaching a crisis in which 350 years of reductionistic science will give way not only to emergence, but to an open universe, partially enabled by enabling constraint laws, partially lawless, uniting physics with history.  There may be lawlessness at the quantum-classical interface, making a Theory of Everything that explains by entailment impossible.  This one universe may have evolved by abiotic natural selection among an infinite or vast set of laws to be a winning universe, perhaps by maximizing the growth of its own Adjacent Possible, hence its own growth. There might be an approach to the ancient question: Why is there something rather than nothing?  Hawkings Godel and the End of Physics may be only the beginning of a new physics.

\section*{Acknowledgement }
This paper was partially funded by iCORE grant to Kauffman, and a Tekkes grant to Kauffman as a Finnish Distinguished Professor.

\end{document}